\documentclass[aps,prl,twocolumn,groupedaddress,floatfix,showpacs]{revtex4-1}
\usepackage{graphicx}
\usepackage{amsmath}
\usepackage{color}

\begin{document}

\title{Fermi Gases in the Two-Dimensional to Quasi-Two-Dimensional Crossover}

\author{Chingyun~Cheng$^{1,2}$, J.~Kangara$^{1}$, I.~Arakelyan$^{1}$, and J.~E.~Thomas$^{1,*}$}

\affiliation{$^{1}$Department of  Physics, North Carolina State University, Raleigh, NC 27695, USA}
\affiliation{$^{2}$Department of Physics, Duke University, Durham, NC 27708, USA}

\date{\today}

\begin{abstract}
We tune the dimensionality of pancake-shaped strongly-interacting $^6$Li Fermi gas clouds from two-dimensional (2D) to quasi-2D, by  controlling the ratio of the radial Fermi energy $E_F$  to the harmonic oscillator energy $h\nu_z$ in the tightly confined direction.  In the 2D regime, where $E_F<<h\nu_z$, the measured radio frequency resonance spectra are in agreement with 2D-BCS  theory.  In the quasi-2D regime, where $E_F\simeq h\nu_z$, the measured spectra deviate significantly from 2D-BCS theory. For both regimes, the measured cloud radii disagree with 2D-BCS mean field theory, but agree approximately with predictions using a free energy derived from the Bethe-Goldstone equation.
\end{abstract}

\maketitle

Quasi-two-dimensional (quasi-2D) geometries  play important roles in high-temperature superconductors~\cite{RevModPhys.72.969}, layered organic superconductors~\cite{doi:10.1080/00107510110108681}, and semiconductor interfaces~\cite{RevModPhys.62.173}. In high-transition temperature copper oxide and organic superconductors, electrons are confined in a quasi-two-dimensional configuration, creating complex, strongly interacting many-body systems, for which the phase diagrams are not well understood~\cite{Norman08042011}. Enhancement of the critical temperature $T_c$ for the quasi-2D regime, as compared to true 2D regime, has been predicted for thin films in parallel magnetic fields~\cite{Kagan} and for quasi-2D Fermi gases containing atoms in excited states of the tightly confined direction~\cite{PhysRevB.90.214503}, where $T_c$ may exceed the 3D value. Ultracold atomic Fermi gases in 2D and quasi-2D geometries provide model systems, which have been the subject of numerous predictions~\cite{PhysRevLett.62.981,1742-5468-2012-10-P10019, PhysRevB.89.014507,PhysRevB.90.214503,PhysRevLett.95.170407,PhysRevLett.96.040404,PhysRevA.78.033613,PhysRevA.78.043617,
PhysRevA.79.053637,PhysRevLett.106.110403,PhysRevLett.112.135302,PhysRevB.89.014507,Sheehy2DFFLO,
PhysRevA.92.023620,PhysRevA.88.023612,PhysRevB.90.214503,PhysRevA.78.033613} and experiments~\cite{PhysRevLett.105.030404,PhysRevLett.112.045301,PhysRevLett.108.235302,Feld2011, PhysRevLett.108.045302, Koschorreck2012, JochimPairCondensation,Ong2DSpinImbal,PhysRevLett.116.045303,PhysRevLett.116.045302,PhysRevLett.115.010401}.

Experimental studies of interacting Fermi gases in the 2D to quasi-2D crossover are of particular interest, because 2D is the marginal dimension for the formation of confinement-induced quantum bound states~\cite{PetrovShlyapnikov2D} and also for classical fluctuations of the superfluid order parameter. The dimer binding energy $E_b\geq 0$ sets the natural scale of length for scattering interactions in 2D systems~\cite{PhysRevLett.62.981}. As noted by Randeria and Taylor~\cite{doi:10.1146/annurev-conmatphys-031113-133829}, recent experiments have made a number of intriguing and somewhat puzzling observations. Measurements in the nearly 2D regime~\cite{PhysRevLett.108.045302} reveal that the radio frequency absorption threshold is just $E_b$, in agreement with the 2D-BCS mean field prediction~\cite{PhysRevLett.62.981} that one would not have expected to be quantitatively valid in 2D. In contrast, recent experiments on quasi-2D Fermi gases are in disagreement with mean field theory and reveal that a 2D-polaron model can fit both the radio-frequency pairing spectra~\cite{PhysRevLett.108.235302} and the cloud radii for spin-imbalanced and spin-balanced mixtures~\cite{Ong2DSpinImbal}, while it fails to predict the observed phase separation~\cite{Ong2DSpinImbal}. An undamped, monopole breathing mode is found to oscillate at twice the trap frequency for a broad range of temperatures and couplings across the 2D crossover~\cite{PhysRevLett.108.070404}. This apparent scale-invariant behavior, in a theory with an explicit scale $E_b$, is also very surprising~\cite{PhysRevLett.109.135301}, but emerges naturally from zero temperature mean field theory. An important open question is to understand how mean field theory can be applicable for some measurements in the 2D to quasi-2D crossover regime, while it fails for others.

In this Letter, we report measurements of radio frequency spectra and radial cloud profiles for a two-component $^6$Li Fermi gas that is smoothly tunable from the 2D to the quasi-2D regime.  In the 2D regime, we find that the measured spectra  are fit by 2D-BCS mean field theory. For the quasi-2D regime, the measured spectra are fit by a 2D-polaron model, but are inconsistent with 2D-BCS theory. In contrast to the spectra, we find that the measured cloud radii for both the 2D and quasi-2D clouds are inconsistent with 2D-BCS mean field theory, which predicts ideal gas density profiles, but are consistent with the 2D-polaron model. Our results suggest that the success of mean field theory for some measurements may be only apparent.

\begin{figure}[t]
\includegraphics[width=3.0in]{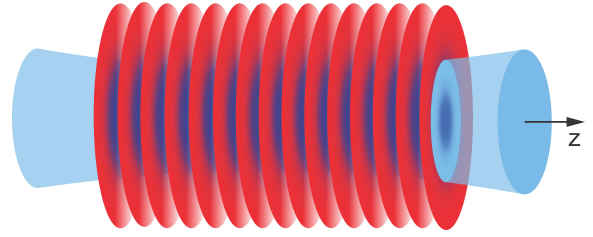}
\caption{The radial confinement of a focused CO$_2$ laser beam (blue) controls the dimensionality of pancake-shaped clouds (red) in a standing-wave optical lattice. The dimensionality of each cloud  is determined by the ratio of the radial Fermi energy $E_{F}$  to the energy level spacing $h\nu_z$ in the tightly confined z-direction of each pancake site. Increasing the CO$_2$ laser intensity tunes the cloud from two-dimensional,  where $E_F/h\nu_z<<1$, to  quasi-2D, where $E_F/h\nu_z\simeq 1$.\label{fig:CO2RedLattice}}
\end{figure}

In the experiments, Fig.~\ref{fig:CO2RedLattice}, two intersecting beams from a fiber laser operating at 1064 nm create an array of pancake-shaped optical traps separated by $0.75\,\mu$m, which tightly confine atoms along the z-axis. Superposed on this periodic array is a focused CO$_2$ laser beam that provides the dominant radial confinement. We trap a balanced (50-50) mixture of atoms in the two lowest hyperfine components (denoted 1,2) of $^6$Li, tuned near the broad Feshbach resonance at 832.2 G~\cite{BartensteinFeshbach,JochimPreciseFeshbach}. By changing the CO$_2$ laser intensity, we vary the ideal 2D gas radial Fermi energy,  $E_F=\hbar\omega_\perp\sqrt{N}$, where $\omega_\perp$ is the radial harmonic oscillator frequency of a noninteracting atom in the trap and $N\sim2000$ is the total number of atoms in one site. The interaction strength is characterized by the parameter $E_F/E_{b12}$, where $E_{b12}$ is the binding energy of a $1-2$ dimer in the pancake trap~\cite{PhysRevLett.108.235302}. We control the dimensionality of each pancake-shaped site by tuning the radial $E_{F}$ relative to the fixed harmonic oscillator energy level spacing $h\nu_z$ in the tightly confined z-direction, while keeping $E_{b12}$ nominally the same. The cloud is 2D for $E_F/h\nu_z<<1$ and quasi-2D for $E_F/h\nu_z\simeq 1$.

\begin{figure}[htb]
\includegraphics[width=3.0in]{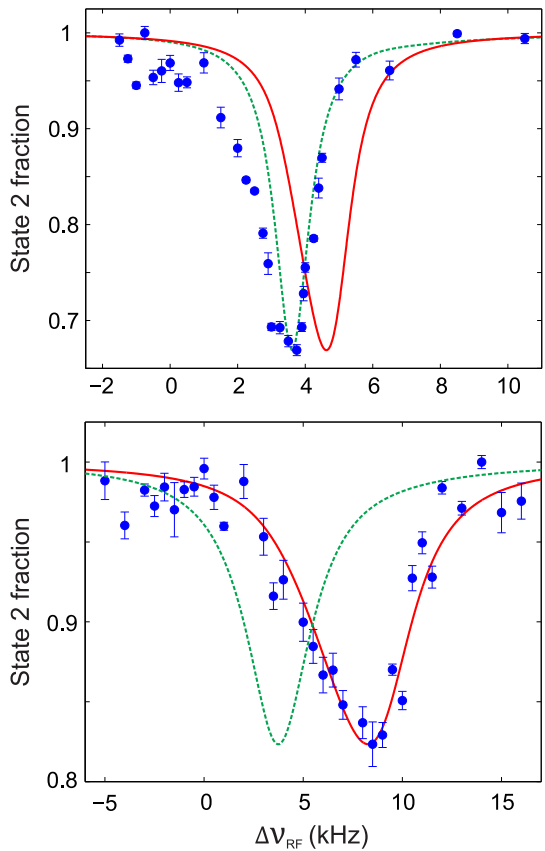}
\caption{Radio-frequency spectra at $B=1005$ G and $\nu_z=98$ kHz. Top: 2D regime with $E_F/h\nu_z=0.18$, $E_{b12}/h\nu_z=0.065$, $E_F/E_{b12}=2.82$. Bottom: Quasi-2D regime with $E_F/h\nu_z=0.89$, $E_{b12}/h\nu_z=0.082$, $E_F/E_{b12}=10.7$. The fraction of atoms remaining in hyperfine state $2$ is measured as a function of radio-frequency relative to the bare atom $2\rightarrow 3$ resonance frequency. The dashed-green (solid-red) curves denote the dimer (polaron) prediction with no free parameters (top) and fitted width $w=4.0$ kHz (bottom). \label{fig:Spectra1002}}
\end{figure}

\begin{figure}[htb]
\includegraphics[width=2.99in]{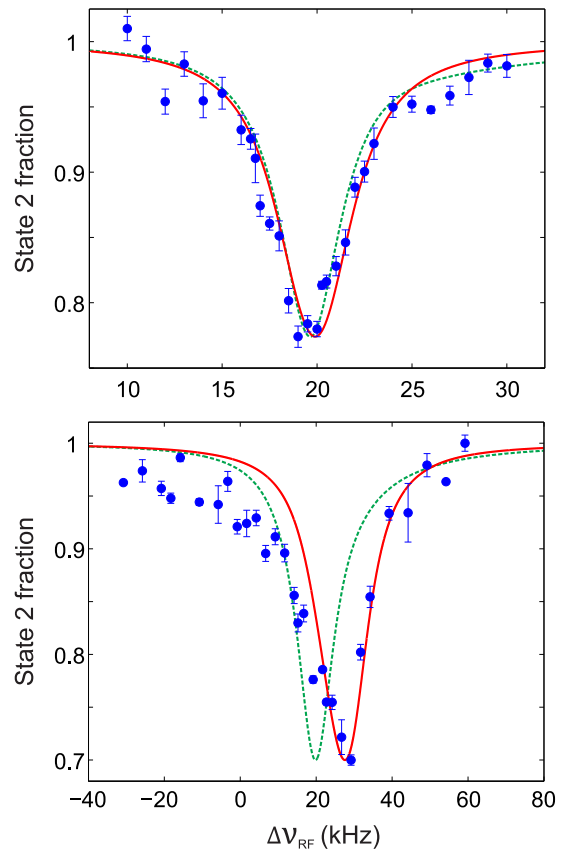}
\caption{Radio-frequency spectra at $B=834$ G and $\nu_z=98$ kHz. Top: 2D regime with $E_F/h\nu_z=0.15$, $E_{b12}/h\nu_z=0.25$, $E_F/E_{b12}=0.60$. Bottom: Quasi-2D regime with $E_F/h\nu_z=0.80$, $E_{b12}/h\nu_z=0.26$, $E_F/E_{b12}=2.99$. The fraction of atoms remaining in hyperfine state $2$ is measured as a function of radio-frequency relative to the bare atom $2\rightarrow 3$ resonance frequency. The dashed-green (solid-red) curves denote the dimer (polaron) prediction with no free parameters (top) and fitted width $w=12$ kHz (bottom). \label{fig:Spectra832}}

\end{figure}

To probe the pairing energy, we use radio-frequency excitation of the transition from the atomic hyperfine state $2$ to a higher lying, initially empty hyperfine state $3$. We record the number of atoms remaining in state 2 as a function of the excitation frequency relative to the bare atom hyperfine transition frequency $\nu^0_{32}$, i.e.,  $\Delta\nu_{RF}\equiv\nu_{rf}-\nu^0_{32}$. We measure $\nu^0_{32}$ using a high temperature, low density $1-2$ mixture, which agrees with measurements for a noninteracting cloud initially containing atoms only in state 2. We then observe the rf spectra in low temperature mixtures, which exhibit a pairing peak, as shown in Fig.~\ref{fig:Spectra1002} for $B=1005$ G and in Fig.~\ref{fig:Spectra832} for $B=834$ G.

We consider first the measurements of rf spectra in the 2D regime. In the simplest picture, the observed location of pairing peaks would arise from the difference between the binding energies of a $1-2$ dimer and a $1-3$ dimer, $h\nu_{rf}=E_{b12}-E_{b13}$, or between a $1-2$ dimer and a $1-3$ scattering state, $h\nu_{rf}=E_{b12}+E_k$ with relative kinetic energy $E_k\geq 0$. However, for the conditions of our experiment, where $E_F\geq E_{b12}$, we expect many-body physics to be important, as the interparticle spacing is then comparable to or smaller than the dimer size.  For the 2D regime, we can try to apply 2D-BCS theory for a true 2D system~\cite{PhysRevA.78.033613}. In this case,  the 2D-BCS prediction for a $2\rightarrow 3$ transition with a noninteracting final state 3 is $h\nu_{rf}=E_{b12}$, which is precisely the dimer pairing energy, as noted previously~\cite{PhysRevLett.108.045302,PhysRevLett.108.235302}. The noninteracting final state approximation is reasonable for our experiments, where $E_{b13}<<E_{b12}$. However, in calculating the spectra, we include $E_{b13}$ in the calculation of the 1-2 dimer to 1-3 dimer transition frequency and we include final state (3-1) interactions in the  threshold spectrum~\cite{PhysRevLett.108.235302}. Further, we convolve the calculated spectra with a Lorentzian of width (FWHM) $w$, which we believe arises from the short lifetime of the excited 3 state in the 1-2 mixture due to three-body collisions. For the data at 1005 G, we use $w=4.8$ kHz, measured using the atomic $2\rightarrow 3$ resonance, which still could be observed in the low temperature $1-2$ mixture. For the data at 834 G, the width $w=15$ kHz is found by fitting, as we could not measure the width of the small atomic resonance signal in this case. For the upper (2D) spectra in Fig.~\ref{fig:Spectra1002} and in Fig.~\ref{fig:Spectra832}, where $E_F/h\nu_z \leq 0.18$, we find that the dimer spectrum predicted by 2D- BCS theory is consistent with the data, as shown by the calculated dashed-green spectra.

Now we examine the measurements in the quasi-2D regime, shown as the lower spectra in Fig.~\ref{fig:Spectra1002} and Fig.~\ref{fig:Spectra832}, where $E_F/h\nu_z \geq 0.8$. Here, we find that 2D-BCS theory does not fit the data. Recently,  zero temperature 2D-BCS theory has been extended to include higher axial states~\cite{PhysRevA.88.023612}, which one expects would contribute in the quasi-2D regime. The predictions show that in the quasi-2D regime, the pairing resonances should be significantly shifted upward in frequency as observed, but quantitative agreement has not yet been obtained.

We consider here a 2D-polaron model, which we have found predicts several features of our previous data in the quasi-2D regime~\cite{PhysRevLett.108.235302,Ong2DSpinImbal}. In the spectra, the polaron model predicts a resonance for $h\nu_{rf}=E_{p13}-E_{p12}$, where the polaron energy of each state is given by
\begin{equation}
E_p=y(q)\epsilon_F.
\label{eq:polaron1}
\end{equation}
Here, $\epsilon_F=\pi\hbar^2\,n/m$ is the local Fermi energy, $m$ is the atom mass, and $n$ is the total density for the 50-50 mixture. An approximate form for the dimensionless factor $y(q)$ has been determined using the Bethe-Goldstone equation~\cite{PhysRevA.84.033607,Klawunn2DEOS} that describes two-body interactions in a many-body system,
\begin{equation}
y(q)=\frac{-2}{ln(1+2\,q)},
\label{eq:polaron2}
\end{equation}
where $q=\epsilon_F/E_b$. This analytic result interpolates between the molecular regime (neglecting the molecular mean field) at magnetic fields well below the Feshbach resonance and agrees with the Fermi polaron approximation~\cite{PhysRevLett.108.235302} and recent QMC predictions~\cite{PhysRevLett.115.185301} at and above the Feshbach resonance, where our data is taken.   The red solid curves  in the spectra of Fig.~\ref{fig:Spectra1002} and Fig.~\ref{fig:Spectra832} show the polaron model predictions
\begin{equation}
I(\Delta\nu)\propto\int \frac{2\pi \rho\,d\rho\,n(\rho)}{1+(2/w)^2\left[\Delta \nu-(E_{p13}-E_{p12})/h\right]^2}.
\end{equation}
The density $n$ and local Fermi energy $\epsilon_F$ decrease with increasing radius from the pancake center, producing a downward sweeping broad spectrum, consistent with the data. We see that the 2D-polaron model explains the quasi-2D data reasonably well, consistent with our previous measurements in a CO$_2$ laser standing wave trap~\cite{PhysRevLett.108.235302}.

In previous studies of quasi-2D  spin-imbalanced and spin-balanced clouds~\cite{Ong2DSpinImbal}, we have measured both the cloud radii and the pressure for $E_F/h\nu_z=1.5$. There, we find that the 2D-polaron model gives a reasonable fit for the measured radii and pressure, while 2D-BCS theory for a balanced gas predicts an ideal gas pressure and  ideal gas cloud profiles~\cite{Ong2DSpinImbal,PhysRevA.78.033613},  in strong disagreement with the measurements. Recently, Fischer and Parish~\cite{PhysRevB.90.214503} have extended finite temperature 2D-BCS theory to include higher axial states, which are expected to contribute to the thermodynamics in the quasi-2D regime. In this case, the predicted pressure decreases below the ideal gas pressure with increasing $E_F/h\nu_z$, but it is well above the 2D-polaron prediction~\cite{Ong2DSpinImbal}, which agrees with measurements in the quasi-2D regime~\cite{Ong2DSpinImbal,PhysRevLett.112.045301}.

\begin{figure}[htb]
\includegraphics[width=3.25in]{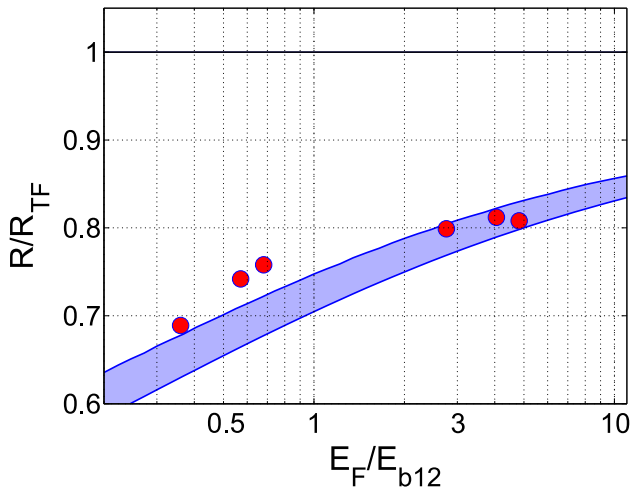}
\caption{Cloud radii versus $E_F/E_b$, where $E_F$ is the radial Fermi energy for an ideal gas, $R_{TF}$ is the Thomas-Fermi radius, and $E_{b12}$ is the 2D dimer binding energy of a $1-2$ atom pair. The blue band shows 2D-polaron model prediction at $T=0$ (lower side) and $T/T_F=0.2$ (upper side). The solid  line at $R/R_{TF}=1$ is the 2D-BCS prediction.\label{fig:CloudRadii2D}}
\end{figure}

Our measured spectra for the 2D regime appear to agree with 2D-BCS mean field theory, which predicts dimer spectra, consistent with the 2D spectra obtained in Ref.~\cite{PhysRevLett.108.045302}. To examine the 2D-BCS predictions further, we use an in-situ phase-contrast method to image the dense clouds in the 2D regime with $E_F/h\nu_z \leq 0.18$. From the atom number and peak column density~\cite{Ong2DSpinImbal}, we obtain the cloud radii shown in Fig.~\ref{fig:CloudRadii2D}.

Over the measured range of $E_F/E_b$, we see that the cloud radii are well below the ideal gas limit $R/R_{TF}=1$, where $R_{TF}=\sqrt{2E_F/m\omega_\perp^2}$ is the Thomas-Fermi radius. In contrast, 2D-BCS theory for a true 2D system predicts ideal gas Thomas-Fermi profiles~\cite{Ong2DSpinImbal,PhysRevA.78.033613}, $R/R_{TF}=1$, in strong disagreement with the data.

Now we consider  a zero-temperature, 2D-polaron model prediction, shown as the the lower side of the blue band in Fig.~\ref{fig:CloudRadii2D}. The cloud radii are determined from the local chemical potential $\mu=\partial f/\partial n$, which is determined from the approximate free energy density for the balanced gas~\cite{Ong2DSpinImbal,Klawunn2DEOS},
\begin{equation}
f=\frac{n}{2}\,\epsilon_F\,[1+y(q)].
\label{eq:freeenergy}
\end{equation}
For the spin-balanced 1-2 mixture, we obtain the cloud radii  in units of the ideal gas Thomas-Fermi radius~\cite{Ong2DSpinImbal},
\begin{equation}
\frac{R}{R_{TF}}=\sqrt{\tilde{\mu}(0)+\frac{E_{b12}}{2E_F}},
\label{eq:radii}
\end{equation}
where $\tilde{\mu}(0)$ is the chemical potential at the center of the cloud in units of $E_F$~\cite{ChemicalPot}.

For these experiments, we are not able to cool the cloud as efficiently as in our previous studies in a CO$_2$ laser lattice, where we obtained $T/T_F<0.2$. We estimate the effect of finite temperature by using ideal gas temperature scaling for the zero temperature radii. For $T/T_F=0.2$, we obtain the upper side of the blue band in Fig.~\ref{fig:CloudRadii2D}.

In conclusion, we have measured both the radio-frequency spectrum and the cloud radii under the same conditions for 2D and quasi-2D Fermi gas clouds. We find that the pairing spectra measured in the 2D regime can be fit using the mean field prediction of dimer spectra, but that the cloud radii in the 2D regime are much smaller than the ideal gas values predicted by 2D-BCS theory. In the quasi-2D regime, quantitatively fitting both the spectra and the cloud radii appears to require a beyond mean field description. From these results, we conclude that the measurement of pairing spectra in the 2D regime~\cite{PhysRevLett.108.045302} does not provide a stringent test of the validity of 2D-BCS theory and that mean-field theory does not quantitatively describe the 2D system, confirming the conjecture by Randeria and Taylor~\cite{doi:10.1146/annurev-conmatphys-031113-133829}.

Primary support for this research is provided by the Division of Materials Science and Engineering, the Office of Basic Energy Sciences, Office of Science, U.S. Department of Energy (DE-SC0008646) and by the Physics Division of the Army Research Office (W911NF-14-1-0628). Additional support for the JETlab atom cooling group has been provided
by the Physics Divisions of the National Science Foundation (PHY-1404135) and the Air Force Office of Scientific Research (FA9550-13-1-0041). \\

$^*$Corresponding author: jethoma7@ncsu.edu

\begin{thebibliography}{40}
\bibitem [{Che()}]{ChemicalPot}%
  \BibitemOpen
  \href@noop {} {}\bibinfo {note} {Here,
  $\tilde{\mu}(0)=\tilde{n}(0)\{1+y[q_0\tilde{n}(0)]+y'[q_0\tilde{n}(0)]/2\}$,
  where $y'(q)\equiv q\,y^2(q)/(1+2q)$ and $q_0=E_F/E_b$. We determine the
  scaled total density $\tilde{n}(0)$ self-consistently from
  $\tilde{n}(0)=1/\sqrt{1+y[q_0\tilde{n}(0)]+y'[q_0\tilde{n}(0)]}$. For a
  detailed explanation, see Ref.~\cite{Ong2DSpinImbal}.}\BibitemShut {Stop}%
\end{thebibliography}
%

\end{document}